\documentclass[floatfix,%
 aip,jcp,
 amsmath,amssymb,
 reprint,
nolongbibliography
]{revtex4-2}

\usepackage[a-1b]{pdfx} 

\usepackage{graphicx}
\usepackage{dcolumn}
\usepackage{bm}

\usepackage[utf8]{inputenc}
\usepackage[T1]{fontenc}
\usepackage{mathptmx}
\usepackage{listings}
\usepackage{rotating} 

\usepackage{color}
\usepackage{dcolumn} 
\usepackage{bm} 
\usepackage{graphicx}
\usepackage{multirow} 
\usepackage{pifont} 
\usepackage{epsfig}
\usepackage{amsmath} 
\usepackage{amssymb}
\usepackage{mathrsfs}
\usepackage{subfigure}
\usepackage{float}
\usepackage{booktabs}
\usepackage{tabularx}
\usepackage{natbib}
\usepackage{gensymb}
\usepackage{rotating}
\usepackage{threeparttable}
\usepackage{comment}
\usepackage{xr}
\usepackage[normalem]{ulem}
\usepackage{physics} 

\begin{document}
  
\author{P. D. Varuna S. Pathirage}
\affiliation{
             Department of Chemistry and Biochemistry,
             Florida State University,
             Tallahassee, FL 32306-4390, USA}   

\author{Stephen H. Yuwono}
\affiliation{
             Department of Chemistry and Biochemistry,
             Florida State University,
             Tallahassee, FL 32306-4390, USA}   

\author{Xiaosong Li}
\affiliation{Department of Chemistry, University of Washington, Seattle, WA 98195, USA}
             
\author{A. Eugene DePrince III}
\email{adeprince@fsu.edu}
\affiliation{
             Department of Chemistry and Biochemistry,
             Florida State University,
             Tallahassee, FL 32306-4390, USA}

\title{Time-Dependent Relativistic Two-Component Equation-of-Motion Coupled-Cluster for Open-Shell Systems: TD-EA/IP-EOMCC}

\begin{abstract}
We present a combined imaginary-time/real-time time-dependent (TD) approach for evaluating linear absorption spectra of open-shell systems at the electron attachment (EA) and ionization potential (IP) equation-of-motion coupled-cluster (EOMCC) levels of theory and within the exact two-component relativistic framework. The absorption lineshape is given by the Fourier transform of the electric dipole autocorrelation function, which is obtained from a real-time simulation. Approximations of the lowest-energy EA- and IP-EOMCC eigenstates, which are required as initial states for the real-time simulation, are generated by propagating a { Koopmans} EA/IP state in imaginary time. TD-EA/IP-EOMCC linear absorption spectra of open-shell atomic {(Na, K, Rb, F, Cl, and Br) and diatomic (SiH and ClO) systems} closely reproduce those obtained from standard TD-EA/IP procedures carried out in the frequency domain. We find that the existence of low-lying states with non-negligible overlap with the { Koopmans} determinant impacts the length of the imaginary-time propagation required to obtain an initial state that produces correct absolute energies and peak height intensities in spectra extracted from the subsequent real-time TD-EA/IP-EOMCC calculations.

\end{abstract}

\maketitle

\section{Introduction}
\label{SEC:INTRODUCTION}

Real-time (RT) time-dependent (TD) electronic structure methods\cite{Li20_review,Pedersen23_e1666} have enjoyed an increase in popularity in recent years because of their usefulness in describing wide-ranging excited-state processes relevant to ultrafast dynamical phenomena,\cite{Stefanucci18_061401, Evangelista19_171102, Cederbaum08_232, Lopata17_3991, Govind11_1344} various spectroscopic signals,\cite{Cramer15_1102,DePrince16_5834,Sheng18_1800055,DePrince17_2951,Bartlett19_164117,Peng20_174113,Peng20_6983,Bartlett21_094103,Rehr22_1799,Repisky23_1714,Li16_234102,Li19_6824,DePrince19_204107} and time-resolved optical experiments.\cite{Mukamel12_194306,Govind15_4294,Rubio16_128,Parkhill16_1590,Lopata20_4470,Koch22_023103,Repisky23_1714} RT-TD approaches also exhibit some desirable practical benefits over frequency-domain representations of excited-state electronic structure. For example, RT-TD methods can be used to obtain broad spectral information in large systems with high densities of states with a significantly reduced memory footprint,\cite{Cramer15_1102,DePrince17_2951} as compared to their frequency-domain counterparts. Frequency-domain calculations built on iterative eigensolver techniques may require the storage of many wave-function-sized objects, whereas RT-TD methods require the storage of only a few such objects. The reduced-storage requirement is particularly desirable for correlated many-body methods such as coupled-cluster (CC)\cite{Coester58_421,Kuemmel60_477,Cizek66_4256,Cizek69_35,Shavitt72_50,Li99_1,Musial07_291} or equation-of-motion (EOM) CC\cite{Emrich81_379,Bartlett89_57,Bartlett93_7029} theory.

In the context of time-independent electronic structure theory,  CC and EOMCC are regarded as ``gold standard'' methods that effectively balance the accurate treatment of electron correlation effects and computational cost. Indeed, the basic CC approach with single and double excitations (CCSD) and its EOMCCSD counterpart can usually provide qualitatively correct descriptions of electronic structure,\cite{Bartlett82_1910,Zerner82_4088,Bartlett93_7029} and the inclusion of higher-order many-body components of the cluster and EOM excitation operators, as is done in CC/EOMCC with up to triple excitations (CCSDT/EOMCCSDT)\cite{Schaefer86_207,Bartlett87_7041,Schaefer88_382,Bartlett90_6104} or CC/EOMCC with up to quadruple excitations (CCSDTQ/EOMCCSDTQ)\cite{Adamowicz91_6645,Bartlett91_387,Bartlett92_4282,Adamowicz94_5792} results in a hierarchy of CC/EOMCC methods that quickly converges to the full configuration interaction (CI) solution.\cite{Musial07_291} Given the successes of CC and EOMCC, it is not surprising that several TD formulations of CC\cite{Klamroth11_054113,Kvaal19_144106,Pedersen20_071102,Koch23_033116,Kvaal12_194109,Ishikawa18_051101,Ishikawa20_124115,Ishikawa20_034110,Ishikawa21_234104,Olsen05_084116} and EOMCC\cite{Schlegel11_4678,Head-gordon12_909,DePrince16_5834,DePrince17_2951,DePrince19_204107,Li19_6617,Bartlett19_164117,Bartlett21_094103,DePrince21_5438,Koch22_023103,Koch23_033116} have been put forward. 

In this work, we focus on the application of moment-based TD-EOMCC approaches\cite{DePrince16_5834, DePrince17_2951, DePrince19_204107, Bartlett19_164117} to the description of open-shell species. The simplest route to spectra of such species would be the direct application of TD-EOMCC to the open-shell states obtained using either unrestricted or restricted open-shell Hartree--Fock reference (abbreviated as UHF and ROHF, respectively). However, it is well known that UHF-based calculations often break total spin ($\hat{S}^2$) symmetry, while ROHF-based calculations are not invariant with respect to the canonicalization of the Fock matrix. Within the EOMCC framework, these problems can be avoided by relying on non-particle-conserving operators characterizing the ionization potential (IP)\cite{Stanton94_65,Snijders92_55,Snijders93_15,Gauss94_8938,Bartlett03_1128,Bartlett04_210,Gauss05_154107,Wloch05_134113,Wloch06_2854,Piecuch06_234107} and electron attachment (EA)\cite{Bartlett95_3629,Bartlett95_6735,Bartlett03_1901,Wloch05_134113,Wloch06_2854,Piecuch06_234107} EOMCC approaches. Unlike in the conventional particle-conserving EOMCC (also known as electron-excitation (EE) EOMCC), IP-EOMCC removes an electron from and EA-EOMCC add an electron to a reference wave function. If this reference is an $\hat{S}^2$ eigenfunction ({\em i.e.},  a restricted Hartree--Fock [RHF] reference), then the resulting open-shell description can maintain rigorous spin adaptation, allowing one to describe radical species with the proper degeneracy structures. To the best of our knowledge, the utility of non-particle-conserving EOMCC {\em ansatze} such as EA/IP-EOMCC in the context of TD-EOMCC remains unexplored.

This paper is organized as follows. In Sec.~\ref{SEC:THEORY}, we lay out the pertinent details of EA/IP-EOMCC and describe the extension of these methods to the time domain. We also discuss the generation of suitable initial states for moment-based TD-EA/IP-EOMCC ({\em i.e.}, approximations to the lowest-energy EA/IP-EOMCC eigenfunctions) via imaginary time propagation, as opposed to solving the ground-state EA/IP-EOMCC eigenvalue problems. Section \ref{SEC:COMPUTATIONAL_DETAILS} then provides the details of the calculations performed in this work, the results of which are presented in Sec.~\ref{SEC:RESULTS}. As test cases, we compute the linear absorption spectra of the group I alkali metals (Na, K, and Rb) and group VII halogen atoms (F, Cl, and Br){, as well as two molecular systems (SH and ClO)}. These systems exhibit non-negligible spin--orbit coupling (SOC) effects, which we describe within the exact two-component (X2C) relativistic framework. We also explore the sensitivity of the spectra to the quality of the initial state generated via the imaginary time propagation. Lastly, Sec.~\ref{SEC:CONCLUSIONS} provides an overall summary of the work and some concluding remarks.

\section{Theory}
\label{SEC:THEORY}

\subsection{The EA/IP-EOMCCSD formalism}
\label{SEC:EAIP_THEORY}

We begin by summarizing the IP- and EA-EOMCC formalism, focusing on the EA/IP-EOMCCSD level of theory. The starting point of these methodologies is the CCSD description of a closed-shell reference state, where the wave function is given by
\begin{equation}
\label{eqn:ccsd_wfn}
    \ket*{\Psi_0^{(N)}} = \exp(\hat{T}) \ket{\Phi}
\end{equation}
Here, $\hat{T}$ is the CCSD cluster operator and $\ket{\Phi}$ is an $N$-electron RHF reference determinant. The cluster operator can be expressed as a many-body expansion, which, for CCSD, is given by
\begin{equation}
\label{eqn:ccsd_cluster_op}
    \hat{T} = \hat{T}_1 + {\hat{T}_2} = t_{a}^{i} \hat{a}^a \hat{a}_i + \frac{1}{4} t_{ab}^{ij} \hat{a}^a \hat{a}^b \hat{a}_j \hat{a}_i.
\end{equation}
In Eq.~\ref{eqn:ccsd_cluster_op}, $t_{a}^{i}$ and $t_{ab}^{ij}$ are the singles (or 1-particle--1-hole, 1p--1h) and doubles (or 2-particle--2-hole, 2p--2h) cluster amplitudes, respectively, and $\hat{a}_p$ ($\hat{a}^p$) is a fermionic annihilation (creation) operator acting on the spin orbital labeled $p$. In this work, we use the usual notation where $i,j,\ldots$ and $a,b,\ldots$ are used for labeling spin orbitals that are occupied and unoccupied, respectively, in $\ket{\Phi}$, while the labels $p,q,\ldots$ are used for spin orbitals with generic occupancy. We also employ the Einstein convention, in which repeated upper and lower indices imply summation. The cluster amplitudes are determined by solving the system of non-linear equations
\begin{equation}
    \label{eqn:ccsd_projective}
    \mel*{\Phi_i^a}{\bar{H}}{\Phi} = \mel*{\Phi_{ij}^{ab}}{\bar{H}}{\Phi} = 0,
\end{equation}
where $\bar{H} = \exp(-\hat{T}) \hat{H} \exp(\hat{T})$ is the CCSD similarity-transformed Hamiltonian and $\ket*{\Phi_i^a}$ and $\ket*{\Phi_{ij}^{ab}}$ are singly and doubly substituted determinants, respectively. Subsequently, we evaluate the ground-state energy as
\begin{equation}
\label{eqn:cc_energy}
    E_{0}^{(N)} = \mel*{\Phi}{\bar{H}}{\Phi}.
\end{equation}

To access the ($N\pm1$)-electron states, we employ non-particle-conserving EOM operators on top of the $N$-electron wave function as
\begin{equation}
    \label{eqn:eomR_wfn}
    \ket*{\Psi_{K}^{(N\pm1)}} = \hat{R}_{K}^{(\pm1)} \ket*{\Psi_0^{(N)}} = \hat{R}_{K}^{(\pm1)} \exp(\hat{T}) \ket{\Phi},
\end{equation}
where the $\hat{R}_{K}^{(\pm1)}$ operator is defined as
\begin{equation}
\label{eqn:eaeomR_operator}
    \hat{R}_{K}^{(+1)} = \hat{R}_{K,1}^{(+1)} + \hat{R}_{K,2}^{(+1)} = r_{a}(K) \hat{a}^a + \frac{1}{2} r_{ab}^{\phantom{a}j}(K) \hat{a}^a \hat{a}^b \hat{a}_j,
\end{equation}
in the case of EA-EOMCCSD, and
\begin{equation}
\label{eqn:ipeomR_operator}
    \hat{R}_{K}^{(-1)} = \hat{R}_{K,1}^{(-1)} + \hat{R}_{K,2}^{(-1)} = r^{i}(K) \hat{a}_i + \frac{1}{2} r_{\phantom{i}b}^{ij}(K) \hat{a}^b \hat{a}_j \hat{a}_i,
\end{equation}
in the case of IP-EOMCCSD. Note that, unlike the cluster operator, the EA-EOM operators are of the $n$p{--}$(n-1)$h excitation type, and the IP-EOM many-body components are of the $(n-1)$p{--}$n$h type. Additionally, because our targets are the spectra of the $(N\pm1)$-electron radical species, we will use $K=0$ to refer to the ground state of the open-shell species, which should not be confused with the CCSD wave function $\ket*{\Psi_0^{(N)}}$. The $(N\pm1)$-electron spectrum could be obtained by solving the eigenvalue problem
\begin{equation}
\label{eqn:eomR_eigenvalue_open}
    \bar{H}_\mathrm{N} \hat{R}_{K}^{(\pm1)}\ket{\Phi} = \omega_{K}^{(N\pm1)}\hat{R}_{K}^{(\pm1)}\ket{\Phi},
\end{equation}
where $\bar{H}_\mathrm{N} = \bar{H} - E_{0}^{(N)}$ and $\omega_{K}^{(N\pm1)} = E_{K}^{(N\pm1)} - E_{0}^{(N)}$ correspond to the EA [$\omega_{K}^{(N+1)}$] and IP [$\omega_{K}^{(N-1)}$] energies relative to the $N$-electron reference state. To avoid confusion, we use the symbol $\omega_K$ without the $(N\pm1)$ superscript to denote excitation energies with respect to the relevant lowest-energy $(N\pm1)$-electron state (\emph{i.e.}, $\omega_K = E_{K}^{(N\pm1)} - E_{0}^{(N\pm1)}$).

The similarity transformation of the Hamiltonian results in a loss of Hermiticity, which means that the left- and right-hand eigenvectors of $\bar{H}$ are not Hermitian conjugates of one another (\emph{i.e.}, $\bra*{\tilde{\Psi}_{K}} \neq \ket*\Psi_{K}^\dagger$). Thus, if properties other than energy are of interest, one must solve the left-hand eigenvalue problem for the left-hand wave functions as well. In the EA/IP-EOMCCSD framework, the left-hand wave function is parameterized as
\begin{equation}
\label{eqn:eomL_wfn}
    \bra*{\tilde{\Psi}_{K}^{(N\pm1)}} = \bra{\Phi} \hat{L}_{K}^{(\pm1)} \exp(-\hat{T}),
\end{equation}
where we have introduced the EOM de-excitation operators
\begin{equation}
\label{eqn:eaeomL_operator}
    \hat{L}_{K}^{(+1)} = \hat{L}_{K,1}^{(+1)} + \hat{L}_{K,2}^{(+1)} = l^{a}(K) \hat{a}_a + \frac{1}{2} l_{\phantom{a}j}^{ab}(K) \hat{a}^j \hat{a}_b \hat{a}_a,
\end{equation}
in the case of EA-EOMCCSD, and
\begin{equation}
\label{eqn:ipeomL_operator}
    \hat{L}_{K}^{(-1)} = \hat{L}_{K,1}^{(-1)} + \hat{L}_{K,2}^{(-1)} = l_{i}(K) \hat{a}^i + \frac{1}{2} l_{ij}^{\phantom{i}b}(K) \hat{a}^i \hat{a}^j \hat{a}_b,
\end{equation}
in the case of IP-EOMCCSD. The left-hand EA/IP-EOM amplitudes are obtained by solving the left-hand eigenvalue problem
\begin{equation}
\label{eqn:eomL_eigenvalue_open}
    \bra{\Phi}\hat{L}_{K}^{(\pm1)} \bar{H}_\mathrm{N} = \omega_{K}^{(N\pm1)}\bra{\Phi}\hat{L}_{K}^{(\pm1)},
\end{equation}
and the left- and right-hand eigenvectors form a biorthonormal set, \emph{i.e.}, they satisfy the relationship
\begin{equation}
\label{eqn:biorthonormality}
    \braket*{\tilde{\Psi}_{K}^{(N\pm1)}}{\Psi_L^{(N\pm1)}} = \mel*{\Phi}{\hat{L}_{K}^{(\pm1)}\hat{R}_L^{(\pm1)}}{\Phi} = \delta_{KL},
\end{equation}
where $\delta_{KL}$ is the Kronecker delta. Properties and transition properties between EA/IP-EOMCCSD states can then be computed as
\begin{equation}
\label{eqn:eom_properties}
    \expval*{\hat{\Omega}}_{KL} = \mel*{\tilde{\Psi}_{K}^{(N\pm1)}}{\hat{\Omega}}{\Psi_{L}^{(N\pm1)}} = \mel*{\Phi}{\hat{L}_{K}^{(\pm1)} \bar{\Omega} \hat{R}_L^{(\pm1)}}{\Phi},
\end{equation}
where $\bar{\Omega} = \exp(-\hat{T}) \hat{\Omega} \exp(\hat{T})$ is the similarity-transformed form of the operator of interest $\hat{\Omega}$.

\subsection{Extension of EA/IP-EOMCCSD to the time-domain}
\label{SEC:TDEAIP_THEORY}
In this subsection, we extend the TD-EOMCC formalism described in Refs.~\citenum{DePrince16_5834, DePrince17_2951, DePrince19_204107, Bartlett19_164117} to the EA/IP-EOMCC domain. Again, we focus on the EA/IP-EOMCCSD levels of theory, but the formalism presented here can be extended to higher levels. Focusing on the linear absorption spectra of radicals, our starting point is the isotropically averaged oscillator strength
\begin{equation}
\label{eqn:td_oscillator_strength}
    f(\omega) = \frac{2}{3} \omega \sum_\alpha \Re{[I_{\alpha}(\omega)]},
\end{equation}
where $\omega$ is a frequency, $\alpha$ is a cartesian component, and $I_{\alpha}(\omega)$ is the lineshape function for a given direction in the cartesian coordinate. The lineshape function is given by the Fourier transformation
\begin{eqnarray}
\label{eqn:td_lineshape}
    I_{\alpha}(\omega) &= \mathcal{F} \left[ e^{-\gamma |t|} {\bigg (} \mel*{\tilde{M}_{\hat{\mu}}^{\alpha,(N\pm1)}} {\hat{U}(t)}{M_{\hat{\mu}}^{\alpha,(N\pm1)}} \right .  \nonumber \\
    &{\left . -\exp(i\omega_0 t)|\langle \mu_\alpha\rangle|^2 \bigg ) \right ]}
\end{eqnarray}
where $\gamma$ is a Lorentzian broadening factor, $t$ is time, and $\hat{U}(t) = e^{i\bar{H}_\mathrm{N} t}$. The EA/IP-EOMCC moment functions entering the Fourier transformation, which serve as the initial states in the propagation, are defined as
\begin{equation}
\label{eqn:momentR}
    \ket*{M_{\hat{\mu}}^{\alpha,(N\pm1)}} = \bar{\mu}_{\alpha} \hat{R}_0^{(\pm1)} \ket{\Phi}
\end{equation}
and
\begin{equation}
\label{eqn:momentL}
    \bra*{\tilde{M}_{\hat{\mu}}^{\alpha,(N\pm1)}} = \bra{\Phi} \hat{L}_0^{(\pm1)} \bar{\mu}_{\alpha},
\end{equation}
in which $\bar{\mu}_{\alpha} = \exp(-\hat{T}) \hat{\mu}_{\alpha} \exp(\hat{T})$ is the similarity-transformed $\alpha$ cartesian component of the dipole moment operator in the length gauge. {The second term in Eq.~\ref{eqn:td_lineshape} represents a zero-frequency component of the lineshape function involving the square of the ground-state electric dipole moment,
\begin{align}
\label{eqn:imag_dip_expectation}
\langle \mu_\alpha \rangle = \bra{\Phi} \hat{L}_0^{(\pm1)}\bar{\mu}_{\alpha}\hat{R}_0^{(\pm1)} \ket{\Phi}
\end{align}
Equation \ref{eqn:td_lineshape} is equivalent to an expression for the lineshape function that includes only the first term, provided that the moment functions are defined in terms of similarity-transformed dipole operators that are normal ordered with respect to the ground-state ({\em i.e.} $\bar{\mu}_{\alpha} \to\bar{\mu}_{\alpha} - \langle \mu_\alpha\rangle$).
This second term in Eq.~\ref{eqn:td_lineshape} can usually be safely ignored because the factor of $\omega$ in Eq.~\ref{eqn:td_oscillator_strength} drives this contribution to zero in the oscillator strength function. However, as we will see in Sec.~\ref{SEC:RESULTS}, this term may become relevant at non-zero frequencies for systems with a permanent electric dipole moment if the real-time propagation is seeded with an approximation to the ground-state wave function.} 

As discussed in Ref.~\citenum{DePrince19_204107}, this moment-based formalism can be generalized for other operators and types of linear spectra, but, in this work, we focus on the electric dipole moment operator relevant to electronic linear absorption. Furthermore, the time propagation operator, $\hat{U}(t)$, can be applied to either the bra or ket moment functions; this formalism allows us to propagate just one of them, which is a considerable savings compared to the need to propagate both left- and right-hand functions as is done in field-driven TD-EOMCC simulations. 

The left- and right-hand moment functions can be expressed using many-body expansions and the operators introduced above to represent the lowest-energy ($K=0$) left- and right-hand EA/IP-EOMCCSD wave functions. In the EA-EOMCCSD case, the right- and left-hand amplitudes are defined as
\begin{align}
\label{eqn:ea_momentR}
    m_{a}(\alpha) & = \mel*{\Phi^{a}} {\bar{\mu}_{\alpha} \hat{R}_0^{(+1)}} {\Phi} \nonumber\\
    m_{ab}^{\phantom{a}j}(\alpha) & = \mel*{\Phi_{\phantom{a}j}^{ab}} {\bar{\mu}_{\alpha} \hat{R}_0^{(+1)}} {\Phi}
\end{align}
and
\begin{align}
\label{eqn:ea_momentL}
    \tilde{m}^{a}(\alpha) & = \mel*{\Phi} {\hat{L}_0^{(+1)}\bar{\mu}_{\alpha}} {\Phi^{a}} \nonumber\\
    \tilde{m}_{\phantom{a}j}^{ab}(\alpha) & = \mel*{\Phi} {\hat{L}_0^{(+1)}\bar{\mu}_{\alpha}} {\Phi_{\phantom{a}j}^{ab}},
\end{align}
respectively. In {Eqs.~\ref{eqn:ea_momentR} and \ref{eqn:ea_momentL}}, $\ket*{\Phi^{a}}$ and $\ket*{\Phi_{\phantom{a}j}^{ab}}$ are 1p and 2p{--}1h determinants, respectively, that span the EA-EOMCCSD excitation space. Similarly, the amplitudes for IP-EOMCCSD moment functions are given by
\begin{align}
\label{eqn:ip_momentR}
    m^{i}(\alpha) & = \mel*{\Phi_{i}} {\bar{\mu}_{\alpha} \hat{R}_0^{(-1)}} {\Phi} \nonumber \\
    m_{\phantom{i}b}^{ij}(\alpha) & = \mel*{\Phi_{ij}^{\phantom{i}b}} {\bar{\mu}_{\alpha} \hat{R}_0^{(-1)}} {\Phi}
\end{align}
and
\begin{align}
\label{eqn:ip_momentL}
    \tilde{m}_{i}(\alpha) & = \mel*{\Phi} {\hat{L}_0^{(-1)}\bar{\mu}_{\alpha}} {\Phi_{i}} \nonumber\\
    \tilde{m}_{ij}^{\phantom{i}b}(\alpha) & = \mel*{\Phi} {\hat{L}_0^{(-1)}\bar{\mu}_{\alpha}} {\Phi_{ij}^{\phantom{i}b}},
\end{align}
in which $\ket*{\Phi_{i}}$ and $\ket*{\Phi_{ij}^{\phantom{i}b}}$ are the 1h and 1p{--}2h determinants, respectively, spanning the IP-EOMCCSD excitation space.
It is worth noting that these amplitudes are practically identical to a typical ``sigma build'' expressions characterizing the EA/IP-EOMCCSD diagonalization procedure, except for the use of $\bar{\mu}$ elements instead of those of $\bar{H}$. Thus, one can use a minimally modified EA/IP-EOMCCSD machinery to prepare the initial $\ket*{M_{\hat{\mu}}^{\alpha,(N\pm1)}}$ and $\bra*{\tilde{M}_{\hat{\mu}}^{\alpha,(N\pm1)}}$ states, and subsequently perform the propagation $\hat{U}(t)\ket*{M_{\hat{\mu}}^{\alpha,(N\pm1)}}$ [or $\bra*{\tilde{M}_{\hat{\mu}}^{\alpha,(N\pm1)}}\hat{U}(t)$] using, for example, a Runge--Kutta-type time integration with the usual EA/IP-EOMCCSD sigma builds combined with the $m$ or $\tilde{m}$ amplitudes.

At this point, we highlight the primary differences between the TD-EA/IP-EOMCC formalism outlined above and its EE predecessor introduced in Ref.~\citenum{DePrince16_5834}. In TD-EE-EOMCC, the initial moment functions are generated by the application of the similarity-transformed dipole operator on the $N$-electron CC ground state and its left-hand lambda counterpart. This structure implies that one only needs to solve for the usual lambda equations after obtaining the converged cluster amplitudes and ground-state CC energy in order to initialize TD-EE-EOMCC simulation. In contrast, within the TD-EA/IP-EOMCC framework, the moment functions cannot be evaluated without knowledge of the ground-state left- and right-hand wave functions of the open-shell system, which correspond to the lowest-energy state in the ($N\pm1$)-electron Hilbert space (or subspace defined by the EA/IP-EOMCC truncation level). A second distinction lies in the fact that the normal-ordered similarity-transformed Hamiltonian that appears in the time evolution operator, $\hat{U}(t) = e^{i\bar{H}_\mathrm{N} t}$, should be defined with respect to the ionized/electron attached state, {\em i.e.}, $\bar{H}_\mathrm{N} = \bar{H} - E_{0}^{(N\pm1)}$. Thus, initializing the TD-EA/IP-EOMCC procedure requires one to (i) solve the $N$-electron CCSD problem and (ii) determine the lowest-energy and left- and right-hand EA/IP-EOMCC, along with the associated energy.

In this work, we propose that one may employ imaginary time propagation to obtain the lowest-energy eigenpair within the $(N\pm1)$-electron Hilbert space. This procedure serves as an alternative to the direct solution of the left- and right-hand EA/IP-EOMCC eigenvalue problems. We begin by choosing a single $(N\pm1)$-electron determinant (\emph{i.e.}, a { Koopmans} state), and evolving this state according to
\begin{equation}
\label{eqn:eomR_imaginary}
    \lim_{\tau \rightarrow \infty} \ket*{\Psi_0^{(N\pm1)}(\tau)} = \lim_{\tau \rightarrow \infty} \exp(-\bar{H}\tau) \ket*{\Phi^{(N\pm1)}}
\end{equation}
and 
\begin{equation}
\label{eqn:eomL_imaginary}
    \lim_{\tau \rightarrow -\infty} \bra*{\tilde{\Psi}_0^{(N\pm1)}(\tau)} = \lim_{\tau \rightarrow -\infty} \bra*{\Phi^{(N\pm1)}} \exp(\bar{H}\tau),
\end{equation}
with $\tau = -it$. Here, $\ket*{\Phi^{(N+1)}}=\ket{\Phi^{a}}$ where $a$ is the lowest unoccupied spin orbital, and $\ket*{\Phi^{(N-1)}}=\ket{\Phi_{i}}$ where $i$ is the highest occupied spin orbital. In practice, one need not propagate to $\tau \rightarrow \infty$; rather, one may stop the simulation once the energy, {$E_{0}^{(N\pm1)}(\tau) =\bra*{\tilde{\Psi}_0^{(N\pm1)}(\tau)}\hat{H}\ket*{\Psi_0^{(N\pm1)}(\tau)}$}, changes by less than some predetermined threshold between successive time steps. Additionally, at every imaginary time step, it is necessary to ensure that binormality (\emph{i.e.}, $\braket*{\tilde{\Psi}_0^{(N\pm1)}}{\Psi_0^{(N\pm1)}} = 1$) is enforced.

{
\subsection{Cost analysis of TD-EA/IP-EOMCCSD}
\label{SEC:TDEAIP_COST}

Finally, let us consider the computational costs of the proposed algorithm in terms of the numbers of occupied and unoccupied spin orbitals ($n_\mathrm{o}$ and $n_\mathrm{u}$, respectively), using the EA/IP-EOMCCSD level of theory as an illustration. As commonly done, we assume that $n_\mathrm{o} < n_\mathrm{u}$, which is valid as one approaches the complete basis set limit. It is also worth noting that in both time- and frequency-domain EA/IP-EOMCCSD calculations, the rate-limiting step is formally still the ground-state CCSD calculation, which is characterized by $\mathcal{O}(n_\mathrm{o}^{2} n_\mathrm{u}^{4})$ iterative steps associated with the well-known particle--particle ladder (PPL) contraction, but in the analysis below we focus on the EA/IP steps, as the real-time simulation portion of the algorithm will likely dominate the computational cost, due to the large number of time-propagation steps.}

{In the frequency domain, an EA-EOMCCSD calculation the floating-point cost of the sigma build is $\mathcal{O}(n_\mathrm{o} n_\mathrm{u}^{4})$ due to a PPL-like contraction involving $\bar{H}_\mathrm{N}$ and $\hat{R}_{K,2}^{(+1)}$/$\hat{L}_{K,2}^{(+1)}$. In IP-EOMCCSD, the most expensive term scales as $\mathcal{O}(n_\mathrm{o}^{3} n_\mathrm{u}^{2})$ and involves a ring-like contraction between $\bar{H}_\mathrm{N}$ and $\hat{R}_{K,2}^{(-1)}$/$\hat{L}_{K,2}^{(-1)}$. In a Davidson diagonalization procedure, one would have to store $\hat{R}_K^{(\pm1)}$ amplitudes of the sizes $n_\mathrm{u} + n_\mathrm{o} n_\mathrm{u}^{2}$ (in EA-EOMCCSD) or $n_\mathrm{o} + n_\mathrm{o}^{2} n_\mathrm{u}$ (in IP-EOMCCSD) multiplied by the number of requested roots, with additional memory required to store  the sigma vectors and subspace expansion vectors. Note that the storage requirement is in practice doubled due to the need to store the $\hat{L}_K^{(\pm1)}$ amplitudes as well, although the storage of intermediate and sigma vectors can be reused for both right- and left-hand Davidson iterations. 

The TD-EA/IP-EOMCCSD algorithms retain the same formal floating-point scaling as their frequency-domain counterparts, but the number of time steps required to obtain a well-resolved spectrum can be exorbitant. On the other hand, the storage requirements of TD-EA/IP-EOMCCSD are greatly reduced compared to those of frequency-domain calculations. For example, let us consider the standard 4th-order Runge--Kutta (RK4) time-integrator in both the imaginary- and real-time propagations. Assuming that we already have the $\hat{R}_K^{(\pm1)}$ and $\hat{L}_K^{(\pm1)}$ amplitudes at $t=0$, we just need to form the $\ket*{M_{\hat{\mu}}^{\alpha,(N\pm1)}}$ and $\bra*{\tilde{M}_{\hat{\mu}}^{\alpha,(N\pm1)}}$ states, which is characterized by $\mathcal{O}(n_\mathrm{o} n_\mathrm{u}^3)$ and $\mathcal{O}(n_\mathrm{o}^2 n_\mathrm{u}^2)$ floating-point costs in the case of EA-EOMCCSD and IP-EOMCCSD, respectively. These costs are less than that of a standard sigma build involving the Hamiltonian because the dipole moment $\hat{\mu}$ is only a one-body operator. In the RK4 algorithm for the subsequent real-time propagation, one needs to store a copy of the $\ket*{M_{\hat{\mu}}^{\alpha,(N\pm1)}}$ or ($\bra*{\tilde{M}_{\hat{\mu}}^{\alpha,(N\pm1)}}$) amplitudes at each time $t$ and $t+\dd t$, where $\dd t$ is the time step size, plus four additional containers the size of the amplitudes. As mentioned above, the moment-based TD-EOMCC calculation described here has the advantage of requiring only propagating either the bra or ket states, which means that at any given time we need a total of 7 amplitude-sized containers (6 for the propagated state and one for the conjugate state). If imaginary-time refinement of the Koopmans state is performed, the floating-point cost again has the same scaling as the underlying EA/IP-EOMCCSD sigma builds, but one has to propagate both the $\hat{R}_K^{(\pm1)}$ and $\hat{L}_K^{(\pm1)}$ EOM amplitudes. These left- and right-hand states can be evolved separately, so the storage requirements are roughly the same as for the real-time time evolution algorithm. 
}

\section{Computational Details}
\label{SEC:COMPUTATIONAL_DETAILS}

All CCSD, frequency-domain EA/IP-EOMCCSD, and TD-EA/IP-EOMCCSD calculations were carried out using an in-house Python code developed with the assistance of the \texttt{p$\dagger$q} automated code generation package.\cite{DePrince21_e1954709,DePrince25_6679} The relativistic treatment used in this work was the { molecular mean-field X2C (mmfX2C)}\cite{Ilias09_124116, Visscher14_041107, SeveroPereiraGomes18_174113, Li24_3408} approach; { mmfX2C} Fock matrices were obtained from the Chronus Quantum\cite{Li20_e1436} electronic structure package.
Non-relativistic two-electron integrals were obtained from the \textsc{Psi4}\cite{Sherrill20_184108} quantum chemistry package. The Fock matrices and two-electron integral tensors were transformed to the molecular spinor basis using spinor coefficient matrices obtained from Chronus { Quantum. The} correctness of the resulting code and workflow were verified numerically against the { mmfX2C}-based EA/IP-EOMCCSD implementations in Chronus Quantum\cite{Li20_e1436} and the RHF-based non-relativistic EA/IP-EOMCCSD code in GAMESS.\cite{Musial02_71, Wloch05_134113, Piecuch06_234107, Gordon23_7031} All calculations were carried out using the x2c-SVPall-2c\cite{Weigend17_3696} basis sets obtained from Basis Set Exchange website.\cite{Windus19_4814}

Group I alkali metals (Na, K, and Rb atoms) {and SiH} were treated at the EA-EOMCCSD level of theory, starting from the relevant cation references (Na$^{+}$, K$^{+}$, Rb$^{+}${, and SiH$^{+}$}). Group VII halogen species (F, Cl, and Br atoms) {and ClO} were treated at the IP-EOMCCSD level of theory, starting from the relevant anion references (F$^{-}$, Cl$^{-}$, Br$^{-}${, and ClO$^{-}$}). All of the cation and anion reference systems are closed-shell species, so EA/IP-EOMCCSD yield atomic spectra that recover the correct degeneracy structures (\emph{i.e.}, they correspond to spin-adapted results in the non-relativistic limit).

{Using Chronus Quantum, we performed additional calculations for F using IP-EOMCCSD with x2c-TZVPPall-2c and x2c-QZVPPall-2c\cite{Weigend20_5658} basis sets, IP-EOMCCSD(2p--3h)\cite{Wloch05_134113,Wloch06_2854} with all of the x2c series, and IP-EOMCCSDT\cite{Bartlett03_1128,Bartlett04_210} with x2c-SVPall-2c to analyze the quality of its frequency-domain spectrum. We also computed the IP-EOMCCSD/x2c-SVPall-2c spectrum for Na out of the Na$^{-}$ closed-shell reference to analyze the quality of the EA-EOMCCSD spectrum for Na out of Na$^{+}$. The results of these calculations are reported in the Supporting Information.
}

CCSD calculations were carried out in the molecular spinor basis. Frequency-domain EA/IP-EOMCCSD calculations were carried out by constructing and diagonalizing the full $\bar{H}$ matrix in the space of 1p and 2p--1h excitations (for EA-EOMCCSD) or 1h and 1p--2h excitations (for IP-EOMCCSD). We correlated all occupied and virtual orbitals for the Na$^{+}$ and F$^{-}$ reference systems. For the K$^{+}$ and Cl$^{-}$ reference ions, we used the frozen-core approximation with the innermost core spin orbitals frozen (1s). For the Rb$^{+}$ and Br$^{-}$ reference species, we froze ten core spin orbitals (the [Ne] core). In the case of Rb$^{+}$, the 40 highest-energy virtual orbitals were also frozen. { For both SiH$^{+}$ and ClO$^{-}$, we froze the two lowest-energy occupied spin orbitals (the 1s spin orbitals of Si and Cl).}

For TD simulations, we generated initial states for real-time propagation via imaginary-time evolution of { Koopmans} EA/IP states, using a time step of $\dd{\tau}=0.01$ a.u.~and a total simulation time of up to $\tau$ = 10 a.u. Of course, the real-time simulations could be carried out in the spinor Slater determinant basis, but we chose the diagonal $\bar{H}$ representation in order to use an exact time integration scheme and avoid potential artifacts stemming from approximate numerical integration. The real-time TD simulations used a time step of $\dd{t}=0.01$ a.u.~and a total propagation time of 10000 a.u.~($\approx 242$ fs). For the Fourier transformation of the dipole autocorrelation function [cf.~Eq.~\ref{eqn:td_lineshape}], we used a Lorentzian damping factor of $\gamma=5\times10^{-4}$ a.u., which corresponds to full-width at half maximum (FWHM) of 27.2 meV. We also padded the time-domain signal with zeroes to obtain a total of $2^{23}$ points, resulting in frequency-domain spectra with $\dd \omega = 7.5\times10^{-5}$ $E_\text{h}$~($\approx2$ meV). In order to resolve the SOC-induced splitting of peaks, we performed additional longer real-time TD-EA/IP-EOMCCSD simulations with a total propagation time of 671000 a.u.~(about 16.23 ps) with the same time step of $\dd{t}=0.01$ a.u., and applied a damping factor of $\gamma=1.5\times10^{-5}$ a.u.~(\emph{i.e.}, a FWHM of 0.816 meV). In order to compare the peak heights the time signal is transformed with Riemann-sum approximation as shown below.
\begin{equation}
F(\omega) = \sum_{k=0}^{N-1}y(t_k)\,e^{-i \omega t_k}\,\Delta t,
\end{equation}
where $y(t_n)$ is the time signal at time $t_n$ and $\Delta t =t_{n+1} - t_n$.

\section{Results and Discussion}
\label{SEC:RESULTS}

\subsection{{Alkali metal and halogen atoms}}
\label{SEC:ATOMS}

\begin{figure}[!htpb]
    \includegraphics[width=\linewidth]{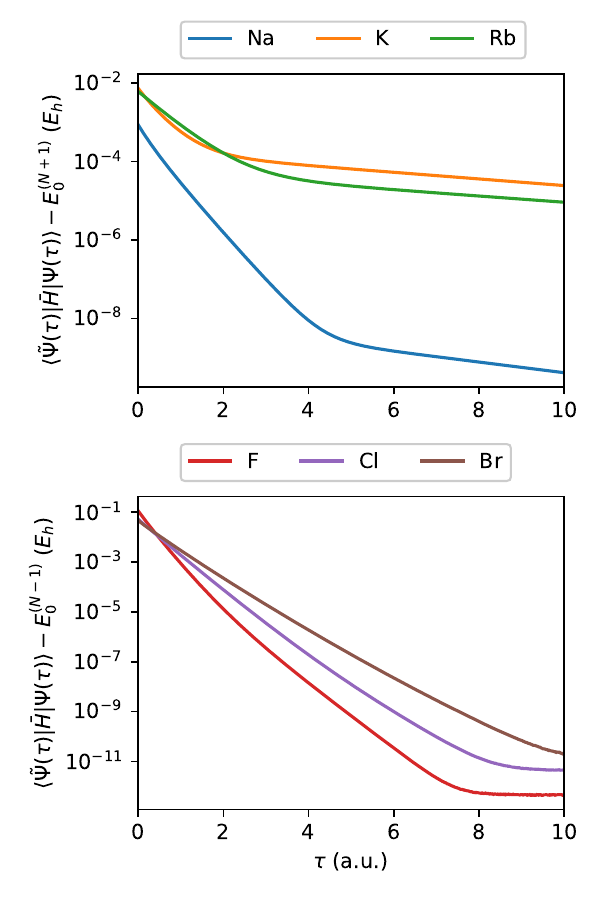}
    \caption{Energy convergence for group I atoms with EA-EOMCCSD (top) and group VII atoms with IP-EOMCCSD (bottom).}
    \label{fig:Imag-prop}
\end{figure}

\begin{table*}[!htbp]
    \caption{Absolute values of the overlap between { Koopmans} states and target EA/IP-EOMCCSD ground states (0) and between the { Koopmans} states and the next lowest state with the same symmetry as the ground state ($K$), along with the excitation energy to state $K$ ($\omega_{K}$) in eV. For each set of degenerate states, only the largest overlap is shown.}
    \begin{tabular*}{\linewidth}{@{\extracolsep{\fill}}lccccc}
        \hline    
        \hline    
        Atom & $|\braket*{\Phi^{(N\pm1)}}{\Psi_0^{(N\pm1)}}|$ & $|\braket*{\tilde{\Psi}_0^{(N\pm1)}}{\Phi^{(N\pm1)}}|$
        & $|\braket*{\Phi^{(N\pm1)}}{\Psi_K^{(N\pm1)}}|$ & $|\braket*{\tilde{\Psi}_K^{(N\pm1)}}{\Phi^{(N\pm1)}}|$ & $\omega_{K}$ \\
        \hline 
        Na & { 0.9998} & { 0.9998} & { 0.0003} & { 0.0002} &  { 4.2448} \\
        K  & { 0.9967} & { 0.9965} & { 0.0414} & { 0.0415} &  { 2.6440} \\
        Rb & { 0.9969} & { 0.9963} & { 0.0250} & { 0.0247} &  { 2.4721} \\
        \hline
        F  & { 0.7040} & { 0.6784} & { 0.0003} & { 0.0003} & { 38.9266} \\
        Cl & { 0.7578} & { 0.7619} & { 0.0004} & { 0.0002} & { 28.9941} \\
        Br & { 0.8146} & { 0.8150} & { 0.0033} & { 0.0035} & { 23.2585} \\
        \hline
        \hline
    \end{tabular*}
    \label{tab:overlap}
\end{table*}

Recall from Eqs.~\ref{eqn:momentR} and \ref{eqn:momentL} that the initial moment functions for TD-EA/IP-EOMCC simulations require knowledge of the ground-state right-hand and left-hand eigenfunctions in the ionized/electron-attached excitation manifolds. These states could be determined as the lowest-energy solutions to the right-hand and left-hand EA/IP-EOMCC eigenvalue problems (Eqs.~\ref{eqn:eomR_eigenvalue_open} and \ref{eqn:eomL_eigenvalue_open}, respectively). On the other hand, as noted in Sec.~\ref{SEC:THEORY}, one could obtain approximations to these states via imaginary time propagation. A question then emerges: how does the convergence of these approximate initial states impact spectra obtained from the subsequent real-time simulations? The accuracy of simulated linear absorption spectra corresponding to approximate initial states could be quantified in three ways: (i) absolute peak positions, (ii) relative peak positions, and (iii) relative peak intensities. The quality of the absolute peak positions will depend on the degree to which the energy associated with $|\Psi^{(N\pm 1)}_0(\tau)\rangle$ agrees with the true ground-state energy of the $(N\pm 1)$-electron states, which is dictated by the energy convergence achieved in the imaginary-time simulation used for state preparation because $E_0^{(N\pm 1)}$ enters the propagator for the real-time simulation through $\bar{H}_N = \bar{H} - E_0^{(N\pm 1)}$. On the other hand, relative positions of the spectral features among themselves will be independent of $\tau$ because $E_0^{(N\pm1)}$ in the propagator serves only to define zero in the excitation spectrum and does not impact relative excitation energies.  Like the absolute peak positions, relative peak intensities will also depend on { the quality} of the initial state.

We can address the accuracy in absolute peak positions by assessing the convergence of the energy associated with the approximate EA/IP-EOMCC ground state, $|\Psi^{(N\pm 1)}_0(\tau)\rangle$, as a function of the imaginary time parameter, $\tau$. Figure~\ref{fig:Imag-prop} depicts the difference between the energy associated with $|\Psi^{(N\pm 1)}_0(\tau)\rangle$ and that of the relevant ground-state energy of the $(N\pm 1)$-electron systems. Data are provided for the group I alkali metal atoms in the top panel (computed via EA-EOMCCSD) and group VII halogen atoms in the bottom panel (computed via IP-EOMCCSD). For the halogen atoms, we observe rapid convergence of the energy in the first 10 a.u.~of imaginary time; at $\tau$ = 10 a.u., the energy error in the bottom panel of Fig.~\ref{fig:Imag-prop} is $\approx$ 10$^{-10}$ $E_\text{h}$, or smaller. The story is slightly different for the alkali metal atoms. For the Na atom, we observe rapid convergence until $\tau$ $\approx$ 4 a.u., followed by slower convergence behavior; by $\tau = 10$ a.u., the energy is converged to less than 10$^{-9}$ a.u. The energy associated with the approximate ground states of the K and Rb atoms converges much more slowly; at $\tau$ = 10 a.u., the energy is only accurate to about 10$^{-5}$ $E_\text{h}$. Data in the Supporting Information show that the energy continues to converge for these systems, over several tens of a.u. By roughly $\tau = 30.0$ a.u., the energies associated with $|\Psi_0^{(N-1)}(\tau)\rangle$ for K and Rb are converged to within a microhartree of the respective ground-state energies. As for the quality of { the} resulting spectra of the $(N\pm1)$-electron states, we can expect errors in absolute peak positions to mirror the errors depicted in Fig.~\ref{fig:Imag-prop}; the largest of these errors at $\tau = 10.0$ a.u.~are on the order of 10$^{-5}$ $E_\text{h}$ or $\approx$ 0.3 meV, which is well within the intrinsic error associated with the EA/IP-EOMCCSD level of theory.

The convergence behavior depicted in Fig.~\ref{fig:Imag-prop} can be understood from the degree to which the initial { Koopmans} state and various open-shell eigenstates of $\bar{H}$ overlap, as well as the proximity of these eigenstates to one another in terms of energy. Table~\ref{tab:overlap} provides the overlap of the { Koopmans} state and the ground-state of $\bar{H}$, as well as the overlap between the { Koopmans} state and the lowest-lying excited state with the same symmetry as the ground state ($^{2}S_{1/2}$ and $^2P_{3/2}$ for the alkali metal and halogen atoms, respectively). The rapid convergence behavior for the halogen atoms is easily rationalized by (i) the small overlap between the { Koopmans} state and the lowest-energy excited $^2P_{3/2}$ symmetry state and (ii) the large energy gap between the this state and the ground state ($\approx$ 23--39 eV). We can see that the rate of convergence follows the degree of overlap in the sense that a large overlap with the excited state leads to slower convergence. On the other hand, for the alkali metal atoms, the lowest-lying $^{2}S_{1/2}$ symmetry states are much closer in energy to the ground state ($\approx 2.5$--4.2 eV), and the degree of overlap with the { Koopmans} states is significantly larger in the cases of K and Rb. As for the halogen atoms, the rate of convergence correlates with the overlap, with Na displaying the most rapid convergence, followed by Rb and K, in that order, with much slower convergence. 

\begin{figure*}[!htbp]
\centering
\includegraphics[width=\linewidth, height=0.85\textheight, keepaspectratio]{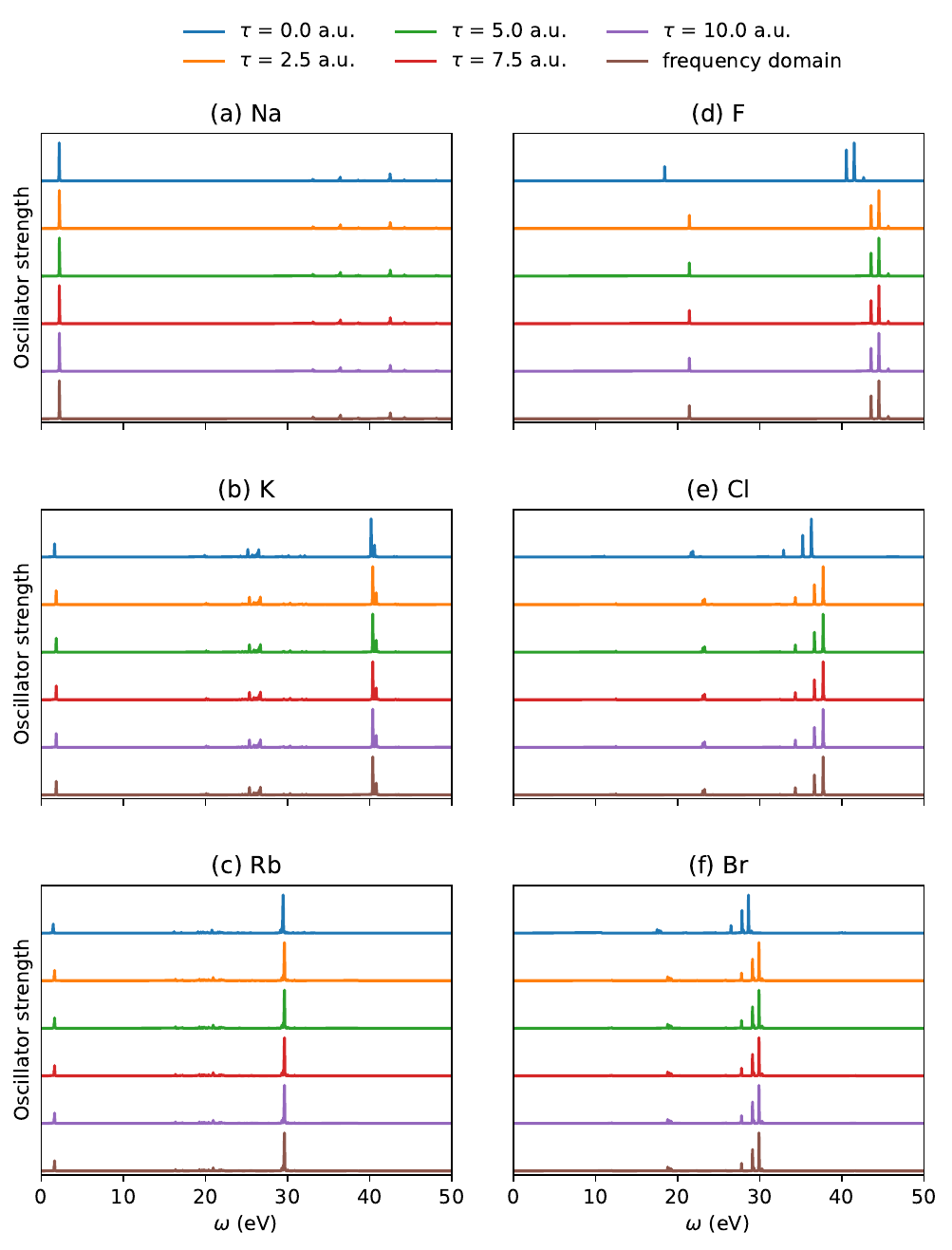}
\caption{{TD-EA/IP-EOMCCSD spectra obtained from simulations seeded with a Koopmans state ($\tau = 0.0$, in blue) or more refined approximations to the ground state ($\tau = 2.5, 5.0, 7.5$ and $10.0$; in orange, green, red, and purple respectively), as well as spectra from standard EA/IP-EOMCCSD calculations} (in brown) for alkali metal and halogen {atoms}. The peak heights are normalized to the highest intensity peak within the range 0 -- 50 eV. For all atoms except for F, the peak with the highest intensity is situated within the range from 0 -- 50 eV. The peak with the highest intensity for F is situated $\approx 107.5$ eV (see Fig.~S2 in Supporting Information).} 
\label{fig:spectra}
\end{figure*}

Figure \ref{fig:spectra} illustrates spectra for the alkali metal and halogen atoms derived from real-time TD-EA/IP-EOMCCSD simulations, with initial states prepared by imaginary-time simulations of various durations, as well as Lorentzian-broadened oscillator strengths taken from standard, frequency-domain EA/IP-EOMCCSD calculations. The spectra are normalized such that the most-intense feature in the energy range depicted has an intensity of one; a similar figure considering a wider energy range can be found in the Supporting Information. We can see that all  features visible in Fig.~\ref{fig:spectra} blue shift with increasing imaginary-time simulation length, $\tau$. This behavior reflects the convergence of the ground-state energy, $E_0^{(N\pm 1)}$, that enters the propagator for the real-time simulation. Consistent with the data in Fig.~\ref{fig:Imag-prop}, the absolute peak positions are visually well-converged within $\tau = 10.0$ a.u. As mentioned earlier, the approximate nature of $E_0^{(N\pm 1)}$ does not impact the relative peak positions, and the data presented in Fig.~\ref{fig:spectra} are in line with this expectation. While relative peak intensities could be sensitive to the quality of the initial state preparation, it is difficult to discern any meaningful errors in peak heights from Fig.~\ref{fig:spectra} at the scale of the figure. 

We assess the impact of initial state quality on relative peak intensities, focusing on zero-field splitting features in these spectra. For the alkali metals, we consider the $^{2}S_{1/2}[ns^{1}]\rightarrow ^{2}P_{1/2},^{2}P_{3/2}[np^{1}]$ transitions, whereas for the halogens we consider $^{2}P_{3/2}[np^5]  \rightarrow ^{2}D_{3/2},^{2}D_{5/2}[np^4(n+1)s^1]$. The ratios of the intensity of the lower-energy peak to that of the higher-energy one for each system are reported in Table~\ref{tab:peak_ratios_671k}. The data indicate that even using the { Koopmans} ($\tau = 0.0$ a.u.) state yields reasonable peak intensity ratios. For Na, K, and Rb, the percent errors in the ratios are all less than 2.0\%. We observe larger errors in the peak intensities for the halogen atoms ({7.5\%, 4.6\%, and 2.4\%} error, respectively, for F, Cl, and Br). These errors rapidly decrease with refinement of the initial state. After only 2.5 a.u.~of imaginary time evolution, the percent error in the peak intensity ratios for all species drop to {0.25}\% or lower. We note that some small residual error persists after 10.0 a.u.~of imaginary time evolution (less than { 0.010}\%), which we suspect is simply a numerical artifact. 

{
Before moving on, let us comment on the quality of the EA/IP-EOMCC spectra presented above, using the smallest atomic systems Na and F as examples. For this analysis, we use results from standard frequency-domain calculations, and the relevant data are provided in the Supporting Information.  To begin with, the EA-EOMCCSD/x2c-SVPall-2c data for Na (out of the Na$^{+}$ reference) show 0.1 eV agreement with experimental data for the low-lying states, whereas the agreement degrades for the higher-energy excited states (1--5 eV errors are observed). Interestingly, we  obtain results of similar quality using IP-EOMCCSD/x2c-SVPall-2c out of the closed-shell Na$^{-}$ reference, indicating that the valence orbitals of the Na$^{+}$ and Na$^{-}$ closed shells are reasonably close to those of neutral Na atom. The situation is quite different for the halogen species. In the case of the IP-EOMCCSD/x2c-SVPall-2c results for F using the F$^{-}$ reference, the lowest-energy $^{2}P_{3/2}$ and $^{2}P_{1/2}$ states are correctly captured, but the excited-state ordering is wrong. For example, the lowest-energy excited state of F should be $^{4}P$ ($2p^{4}3s^{1}$), but IP-EOMCCSD/x2c-SVPall-2c predicts the next excited state to be $^{2}S_{1/2}$ ($2s^{1}2p^{6}$) instead. This discrepancy is a consequence of the noble-gas-like HF determinant for F$^{-}$, which results in a $2p$--$3s$ energy gap that is too large compared to the $2s$--$2p$ energy gap. As shown in the Supporting Information, improving the correlation treatment by including $\hat{R}_{K,3}$ (2p--3h excitation) and  $\hat{T}_3$ or employing larger basis sets such as x2c-TZVPPall-2c and x2c-QZVPPall-2c does not correct the state ordering with respect to available experimental data. These observations suggest that the orbital structure of F$^{-}$ is far from that of the neutral F atom, and we can expect similar behavior for the remaining halogen atoms. Regardless, we would like to emphasize that the main objective of this work is to investigate the viability of Koopmans state as the initial state in a real-time TD-EA/IP-EOMCC propagation, along with the convergence of the Koopmans state to the true EA/IP-EOMCC open-shell ground state via imaginary-time refinement.
}

\begin{table}[!htbp]
    \caption{Peak height ratios (lower-energy/higher-energy) characterizing the $^{2}S_{1/2}[ns^{1}]\rightarrow ^{2}P_{1/2},^{2}P_{3/2}[np^{1}]$ transitions in alkali metals and $^{2}P_{3/2}[np^5]  \rightarrow ^{2}D_{3/2},^{2}D_{5/2}[np^4(n+1)s^1]$ transitions in halogens as functions of the imaginary time $\tau$. The $\tau=\infty$ data in each atom correspond to the results obtained from frequency-domain EA/IP-EOMCCSD calculation. Errors in height ratio are reported as percentage relative to the $\tau=\infty$ values. The total real-time propagation time for these calculations is 671000 a.u..}
    \begin{tabular*}{\linewidth}{@{\extracolsep{\fill}}lccc}
        \hline    
        \hline    
        Atom & $\tau$ (a.u.) & Peak height ratio & error in height ratio \\
        \hline 
        \multirow{6}{*}{Na} & 0.0      & { 0.51101} & { 0.0294}\% \\
                            & 2.5      & { 0.51114} & { 0.0049}\% \\
                            & 5.0      & { 0.51114} & { 0.0049}\% \\
                            & 7.5      & { 0.51114} & { 0.0052}\% \\
                            & 10.0     & { 0.51114} & { 0.0054}\% \\
                            & $\infty$ & { 0.51116} & ---      \\
        \hline 
        \multirow{6}{*}{K}  & 0.0      & { 0.50130} & { 0.8675}\% \\
                            & 2.5      & { 0.49734} & { 0.0703}\% \\
                            & 5.0      & { 0.49699} & { 0.0005}\% \\
                            & 7.5      & { 0.49696} & { 0.0049}\% \\
                            & 10.0     & { 0.49696} & { 0.0048}\% \\
                            & $\infty$ & { 0.49699} & ---      \\
        \hline 
        \multirow{6}{*}{Rb} & 0.0      & { 0.48034} & { 1.1997}\% \\
                            & 2.5      & { 0.48558} & { 0.1234}\% \\
                            & 5.0      & { 0.48606} & { 0.0245}\% \\
                            & 7.5      & { 0.48612} & { 0.0109}\% \\
                            & 10.0     & { 0.48614} & { 0.0083}\% \\
                            & $\infty$ & { 0.48618} & ---      \\
        \hline
        \multirow{6}{*}{F}  & 0.0      & { 0.16727} & { 7.5145}\% \\
                            & 2.5      & { 0.18081} & { 0.0265}\% \\
                            & 5.0      & { 0.18087} & { 0.0028}\% \\
                            & 7.5      & { 0.18087} & { 0.0031}\% \\
                            & 10.0     & { 0.18087} & { 0.0031}\% \\
                            & $\infty$ & { 0.18086} & ---      \\
        \hline
        \multirow{6}{*}{Cl} & 0.0      &  { 0.41461} & { 4.5580}\% \\
                            & 2.5      &  { 0.43443} & { 0.0179}\% \\
                            & 5.0      &  { 0.43441} & { 0.0001}\% \\
                            & 7.5      &  { 0.43441} & { 0.0001}\% \\
                            & 10.0     &  { 0.43441} & { 0.0001}\% \\
                            & $\infty$ &  { 0.43441}   & ---      \\
        \hline
        \multirow{6}{*}{Br} & 0.0      &  { 1.40651} & { 2.3604}\% \\
                            & 2.5      &  { 1.43696} & { 0.2466}\% \\
                            & 5.0      &  { 1.44020} & { 0.0216}\% \\
                            & 7.5      &  { 1.44048} & { 0.0021}\% \\
                            & 10.0     &  { 1.44051} & { 0.0002}\% \\
                            & $\infty$ &  { 1.44051}   & ---      \\
        \hline
        \hline
    \end{tabular*}
    \label{tab:peak_ratios_671k}
\end{table}

\subsection{{Diatomic molecules}}
\label{SEC:DIATOMICS}

\begin{figure*}[!htbp]
\centering
\includegraphics[width=\linewidth, height=0.85\textheight, keepaspectratio]{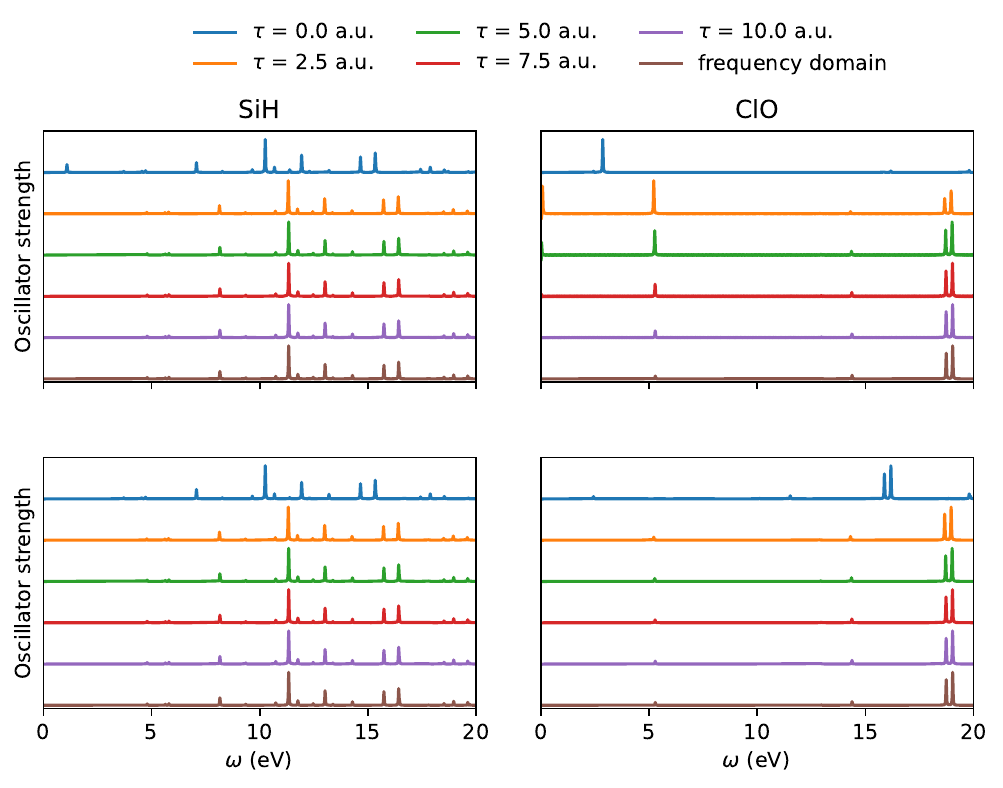}
\caption{{TD-EA/IP-EOMCCSD spectra obtained from simulations seeded with a Koopmans state ($\tau = 0.0$, in blue) or more refined approximations to the ground state ($\tau = 2.5, 5.0, 7.5$ and $10.0$; in orange, green, red and purple respectively), as well as spectra from standard EA/IP-EOMCCSD calculations (in brown) for SiH (left) and ClO (right). The spectra in the top panels were computed with the unmodified dipole moment matrix. The spectra in the bottom panels are computed by subtracting the expectation value of the dipole moment of initial Koopmans or imaginary propagated states as in Eq.~\ref{eqn:imag_dip_expectation} from the similarity transformed dipole moment entering Eqs.~\ref{eqn:momentR} and \ref{eqn:momentL}. In each of the spectra, the peak heights are normalized with respect to the tallest peak within the $\omega=0$--20 eV region. }}
\label{fig:spectra_SiH-ClO}
\end{figure*}

{ We now turn out attention to two diatomic molecular systems: SiH and ClO. As discussed in Sec.~\ref{SEC:ATOMS}, the accuracy of the absolute peak positions in the linear absorption spectra is tied to the quality of the energy expectation value associated with the approximate initial state that seeds the real-time propagation. Figure ~\ref{fig:Imag-prop_SiH-ClO} depicts the convergence of the energy for the initial states as a function of the imaginary time, $\tau$. In the case of SiH, we observe a monotonic decrease in the energy error for the initial state, relative to that of the true $^2\Pi_{1/2}$ ground state computed at the EA-EOMCCSD level. The roughly 0.04 $E_\text{h}$ error at $\tau = 0$ a.u.~decreases to about $2\times10^{-7}$ $E_\text{h}$ by $\tau = 10$ a.u. For ClO, the energy at $\tau = 0.0$ a.u.~is 0.1 $E_\text{h}$ higher than the $^2\Pi_{3/2}$ state energy computed at the IP-EOMCCSD level. The energy error converges rapidly until $\tau$ $\approx$ 1 a.u., after which convergence slows; the energy error is $\approx10^{-4}$ $E_\text{h}$ (2.7 meV) at $\tau = 10$ a.u. Data provided in the Supporting Information show that the correct ground-state energy is recovered to within one microhartree by $\tau = 22.5$ a.u.

\begin{figure}[!htpb]
    \includegraphics[width=\linewidth]{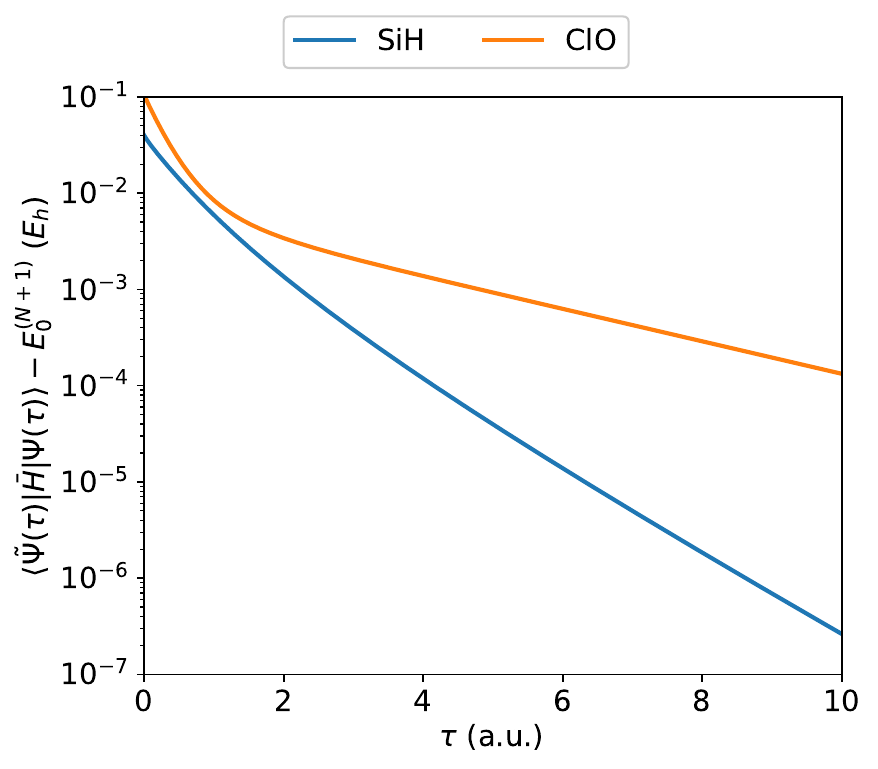}
    \caption{{ Energy convergence for SiH with EA-EOMCCSD (blue) and ClO with IP-EOMCCSD (orange).}}
    \label{fig:Imag-prop_SiH-ClO}
\end{figure}

As was the case for the atoms discussed in Sec.~\ref{SEC:ATOMS}, the convergence behavior shown in Fig.~\ref{fig:Imag-prop_SiH-ClO} is related to the degree of overlap between the initial Koopmans state and the eigenstates of $\bar{H}$. Table~\ref{tab:overlap_SiH-ClO} provides the overlap of the Koopmans state and the ground-state of $\bar{H}$, as well as the overlap between the Koopmans state and the lowest-lying excited state with the same symmetry as the ground state. The steady convergence of SiH can be attributed to the small overlap between the Koopmans state and the lowest-energy excited state at $\approx 2.78$ eV with the same symmetry as Koopmans state. In contrast, the significant overlap between the Koopmans state and the lowest-energy excited state in ClO ($\approx 5.29$ eV) explains the slower convergence behavior.

\begin{table*}[!htbp]
    \caption{{ Absolute values of the overlap between Koopmans states and target EA/IP-EOMCCSD ground states (0) and between the  Koopmans states and the next lowest state with the same symmetry as the ground state ($K$), along with the excitation energy to state $K$ ($\omega_{K}$) in eV, for the diatomic systems. For each set of degenerate states, only the largest overlap is shown.}}
    \begin{tabular*}{\linewidth}{@{\extracolsep{\fill}}lccccc}
        \hline    
        \hline    
        {Molecule} & {$|\braket*{\Phi^{(N\pm1)}}{\Psi_0^{(N\pm1)}}|$} & { $|\braket*{\tilde{\Psi}_0^{(N\pm1)}}{\Phi^{(N\pm1)}}|$}
        & { $|\braket*{\Phi^{(N\pm1)}}{\Psi_K^{(N\pm1)}}|$} & { $|\braket*{\tilde{\Psi}_K^{(N\pm1)}}{\Phi^{(N\pm1)}}|$} & {$\omega_{K}$} \\
        \hline 
        { SiH} & { 0.9797} & { 0.9784} & { 0.0000} & { 0.0002} &  { 2.7771} \\
        \hline
        { ClO}  & { 0.8409} & { 0.8256} & { 0.1157} & { 0.1156} & { 5.2878} \\
        \hline
        \hline
    \end{tabular*}
    \label{tab:overlap_SiH-ClO}
\end{table*}

We now consider linear absorption spectra for SiH and ClO extracted from real-time TD-EA/IP-EOMCCSD simulations in which the initial states at $t=0$ a.u.~are defined by snapshots in the imaginary time evolution used to refine the Koopmans state. These spectra, along with Lorentzian-broadened spectra from standard, frequency-domain EA/IP-EOMCCSD, are depicted in Fig.~\ref{fig:spectra_SiH-ClO}. Two sets of TD-EA/IP-EOMCCSD spectra are provided that differ in the treatment of the ground-state permanent electric dipole contribution to the absorption lineshape function (see Eq.~\ref{eqn:td_lineshape}). As mentioned in Sec.~\ref{SEC:TDEAIP_THEORY}, the permanent electric dipole term can usually be neglected in TD-EOMCC simulations because its contribution at $\omega_0 = 0$ eV is zeroed by the factor of $\omega$ appearing in Eq.~\ref{eqn:td_oscillator_strength}. However, this term cannot be ignored in the present TD-EA/IP-EOMCCSD simulations due to the use of the approximate initial states with $\omega_0 = \langle \tilde{\Psi}(\tau)|\hat{H}|\Psi(\tau)\rangle - E_{0}^{(N\pm1)} \neq 0$ eV. As discussed in Sec.~\ref{SEC:TDEAIP_THEORY}, we could account for this term as expressed in Eq.~\ref{eqn:td_lineshape}, or, alternatively, one could discard the permanent dipole term, provided that the moment functions are defined in terms of similarity-transformed dipole operators that are normal ordered with respect to the ground-state. In TD-EA/IP-EOMCCSD, however, we do not have access to the exact ground state for the $(N\pm 1)$-electron systems, so our only recourse is to normal order with respect to the approximate initial state ({\em i.e.} $\bar{\mu}_{\alpha} \to\bar{\mu}_{\alpha} - \langle \tilde{\Psi}(\tau)|\hat{\mu}_\alpha|\Psi(\tau)\rangle$).

The top panels of Fig.~\ref{fig:spectra_SiH-ClO} depict TD-EA/IP-EOMCCSD spectra generated with unmodified dipole integrals ({\em i.e.}, ignoring the permanent dipole contribution to the lineshape function). For SiH (top left panel), when seeding the real-time simulation with the Koopmans ($\tau = 0$ a.u.) state, we observe a spurious low-energy feature at 1.08 eV, which arises due to the neglect of the permanent dipole term in the lineshape function. This term appears at a non-zero frequency because the Hamiltonian used in time-evolution operator is normal ordered with respect to the the approximate initial state, and the frequency at which this feature appears ($\omega_0$) reflects the error in the energy expectation value associated with the approximate initial state, as compared to the exact ground-state energy of $\bar{H}$. As the initial state is refined with longer simulations in imaginary time, $\omega_0$ approaches zero, and this term vanishes. By $\tau = 2.5$ a.u., we can no longer distinguish the TD-EA-EOMCCSD spectrum from the EA-EOMCCSD one on the scale of this figure. For ClO (top right panel), we observe the same spurious low-energy feature as in SiH (at 2.86 eV), but the peak height is so large that it dwarfs all other features in the $\omega=0$--20 eV range. This feature rapidly red-shifts, but it is still visible on the scale of the figure at $\tau = 7.5$ a.u., reflecting the slow convergence of the energy of the initial state in the imaginary time simulations (see Fig.~\ref{fig:Imag-prop_SiH-ClO}). At $\tau = 2.5$ a.u., the height of the spurious peak has reduced to the point that the other features in the spectrum are discernable, but we can see that the peak height for the first excited-state feature at $\omega\approx2.6$ eV is far to large. The height of this feature slowly converges toward the standard IP-EOMCCSD limit, but the height remains too large compared to the other features, even at $\tau = 10$ a.u.

The bottom panels of Fig.~\ref{fig:spectra_SiH-ClO} depict TD-EA/IP-EOMCCSD spectra generated with dipole integrals that are normal ordered with respect to the approximate initial state, which should partially account for the permanent electric dipole contribution to the lineshape function. Indeed, for both SiH (bottom left panel) and ClO (bottom right panel), shifting the dipole operator eliminates the spurious features at $\omega_0$, even when using the Koopmans state to seed the real-time simulations. Moreover, for ClO, using the normal-ordered dipole operator fixes the peak-height issues discussed above. The TD-IP-EOMCCSD spectrum closely resembles the IP-EOMCCSD spectrum after refining the initial state for only $\tau = 2.5$ a.u.}

\section{Conclusions}
\label{SEC:CONCLUSIONS}

We have presented a time-dependent extension of the EA/IP-EOMCC theory of open-shell systems, incorporating relativistic effects via the X2C formalism. This TD-EA/IP-EOMCC framework enables direct access to linear absorption spectra without the need for iterative diagonalization of the similarity-transformed Hamiltonian as an alternative route to frequency-domain EOMCC calculations.

We have shown that a { Koopmans} determinant constructed by adding or removing a single electron from a closed-shell reference can serve as an effective starting point for real-time propagation for open-shell atomic systems (as opposed to the ``exact'' initial state, which is the lowest-energy eigenfunction of $\bar{H}$ in the $(N\pm1)$-electron manifold). The primary difference between spectra resulting from the use of the { Koopmans} initial state and the frequency-domain EA/IP-EOMCCSD results is a uniform shift, which rapidly decreases when the initial state is refined via evolution in imaginary time. The imaginary time propagation is a numerically inexpensive and conceptually straightforward alternative to traditional Davidson diagonalization for obtain the exact initial state. Like the uniform shift, we have shown that peak intensity ratios associated with zero-field splitting features also approach the correct values as the initial state is refined. 

\vspace{0.5cm}

{\bf Supporting Information} {
Eight CSV files containing the list of transition frequencies and oscillator strenghts obtained from frequency-domain EA/IP-EOMCCSD calculations reported in this work, comparison of EA- and IP-EOMCCSD spectra for the Na atom, comparison of IP-EOMCC spectra for the F atom obtained at different levels of theory and basis sets, energy} convergence for imaginary time propagation of alkali metal atoms up to $\tau = 100.0$ a.u., broad-band TD-EA/IP-EOMCCSD linear absorption spectra, overlaps between approximate initial states at various $\tau$ and eigenstates of $\bar{H}$, peak intensity ratios for zero-field splitting features{, and convergence for imaginary time propagation of ClO up to $\tau = 100.0$ a.u..}

\vspace{0.5cm}

\begin{acknowledgments}
This material is based upon work supported by the National Science Foundation under Grant No. OAC-2103705. SHY acknowledges funding from the Quantum Initiative at Florida State University. \\ 
\end{acknowledgments}

\bibliography{main}

\end{document}